# Performance Analysis of a Ge/Si Core/Shell Nanowire Field Effect Transistor

*Gengchiau Liang,[†,*] Jie Xiang,[‡] Neerav Kharche,[†] Gerhard Klimeck,[†] Charles M. Lieber,[‡,#] and Mark Lundstrom[†]*

[†]School of Electrical and Computer Engineering and Network for Computational Nanotechnology, Purdue University, West Lafayette, IN 47907

[‡]Department of Chemistry and Chemical Biology, [#]Division of Engineering and Applied Sciences, Harvard University, Cambridge, MA 02138, USA

**Abstract**

We analyze the performance of a recently reported Ge/Si core/shell nanowire transistor using a semiclassical, ballistic transport model and an $sp^3s^*d^5$ tight-binding treatment of the electronic structure. Comparison of the measured performance of the device with the effects of series resistance removed to the simulated result assuming ballistic transport shows that the experimental device operates between 60 to 85% of the ballistic limit. For this ~15 nm diameter Ge nanowire, we also find that 14-18 modes are occupied at room temperature under ON-current conditions with $I_{ON}/I_{OFF}=100$. To observe true one dimensional transport in a <110> Ge nanowire transistor, the nanowire diameter would have to be much less than about 5 nm. The methodology described here should prove useful for analyzing and comparing on common basis nanowire transistors of various materials and structures.

[*] Corresponding author Email: liangg@purdue.edu

Semiconducting nanowire transistors are attracting attention due to their potential applications such as electronics[1-7] and biomolecule detection[8,9]. Promising device performance has recently been reported for Si[2,6] and Ge[1] nanowire field-effect transistors (NWFETs). High hole/electron mobilities, large ON-currents, large $I_{ON}/I_{OFF}$ ratios, and good subthreshold swings have been reported[1,2,5,6]. These device performance metrics provide important measures of progress as device fabrication technologies are being refined, but it is still unclear how measured results compare against theoretical expectations, how to compare the results from different experiments, and how to assess nanowire transistor performance against that of state-of-the-art silicon metal-oxide-semiconductor FETs (MOSFETs). Mobility is commonly used as a device metric, but it is not a well-defined concept at the nanoscale, and its relevance to nanoscale MOSFETs is unclear. It is, rather, more appropriate to compare a nanoscale MOSFET against its ballistic limit. In this letter, we do so by analyzing the performance of a recently reported Ge/Si core/shell NWFET[1].

We analyze the performance of a NWFET by comparing the measured current vs. voltage (*I-V*) characteristics to a theoretical model of a ballistic nanowire MOSFET. This semiclassical, top-of-the-barrier model requires as inputs the electronic structure of the nanowire and the gate and drain capacitances[10]. To obtain the bandstructure of the Ge nanowire, we assume an unrelaxed nanowire atomic geometry with bulk atomic positions and construct the Hamiltonian of the nanowire unit cell using the orthogonal-basis $sp^3d^5s^*$ tight-binding method developed for bulk electronic structure[11]. Each atom is modeled using 10 orbitals per atom per spin (20 orbitals per atom total). The nanowire is assumed to be infinitely long, and the nanowire surface is taken to be passivated by hydrogen atoms, which is treated numerically using a hydrogen termination model of the $sp^3$ hybridized interface atoms[12]. This technique has been reported to successfully remove all the interface states from the band gap[12]. Although no relaxation or strain effects are included, this model has shown good agreement with the measured bandgap vs. diameter of silicon nanowires[13].

Figures 1(a) and 1(b) display the computed bandstructure for 5 and 15 nm diameter <110> Ge nanowires. Due to quantum confinement, the direct bandgap at the $\Gamma$-point (projection of $L$ valley of bulk Ge bandstructure) of the nanowire is larger than the indirect bandgap of bulk Ge and increases as the diameter of nanowire decreases, as shown in Fig. 1(c). A similar phenomenon in Si nanowires has been theoretically predicted by different studies[14,15]. The results shown in Fig. 1(c) indicate that in order to see significant quantum confinement effects, the diameter of the nanowire should be smaller than about 5 nm.

To simulate the ballistic *I-V* characteristic of the NWFET, a semi-classical top-of-the-barrier MOSFET model was used[10]. In this model, a simplified 3-dimensional self-consistent electrostatic model including quantum capacitance effects is coupled with a ballistic treatment of hole transport. Three dimensional electrostatics are described by a simple capacitance model[10]. The capacitors represent the electrostatic coupling of the gate ($C_G$), drain ($C_D$), and source terminals ($C_S$) to the top of the potential barrier at the source end of the channel. These capacitors control the subthreshold swing, $S$, of the transistor and the drain-induced barrier lowering (*DIBL*) according to

$$\frac{C_G}{C_\Sigma} = \frac{2.3 k_B T / q}{S} \tag{1a}$$

$$\frac{C_D}{C_\Sigma} = \frac{2.3 k_B T / q}{S} \times DIBL \tag{1b}$$

$$C_\Sigma = C_G + C_D + C_S \tag{1c}$$

The gate insulator capacitance is the most critical parameter in the model. The maximum capacitance would be achieved in a cylindrical gate geometry as shown in Fig. 2(a). For top gated devices, however, a half-cylinder geometry as shown in Fig. 2(b) may be a closer approximation to the actual structure. We used the finite element package, FEMLAB®, to compute the theoretical capacitance for such a structure. In practice, the actual capacitance may be difficult to estimate because of uncertainties in film thicknesses and in the geometry of the gate stack. Measurement of the actual gate capacitance on the actual device being analyzed would be the best procedure, but such measurements are difficult. (Very recently, however, similar measurements on a carbon nanotube transistor have been reported[16].) Accordingly, we will consider both the cylindrical and half cylindrical geometries in the analysis that follows. The uncertainty in gate capacitance is one of the most significant contributions to the error bars for our analysis.

The Poisson's potential ($U_P$) is equal to $U_0 \cdot (N-N_0)$, where $U_0 = q/C_\Sigma$ is the single electron charging energy, $N_0$ and $N$ are the number of mobile carriers at the top of the barrier at equilibrium and under applied bias, respectively, and $C_\Sigma$ is the total capacitance. The carrier density $N$, moreover, can be directly computed from the previously determined $E$-$k$ relations,

$$N = \int_{-\infty}^{\infty} \frac{dk}{\pi} [f(E(k) + U_{scf} - E_{fs}) + f(E(k) + U_{scf} - E_{fs} + qV_D)] , \qquad (2)$$

where $f(E)$ is the Fermi function and $E_{fs}$ is the chemical potential in the source region. Iteration between $N$ and $U_{scf}$ is repeated until the self-consistency reaches convergence. The NW MOSFET current is then evaluated using the semi-classical transport equation in the ballistic limit:

$$I = \frac{2q}{h} \int_{U_{scf}}^{\infty} dE[f(E - E_{fs}) - f(E - E_{fs} + qV_D)] \qquad (3)$$

More details of this model can be found in Ref. 10 and 17.

Using the techniques described above, we analyzed the performance of a recently reported Ge/Si, core/shell nanowire FET[1]. The nominal diameter of the Ge core is $D = 14.7 \pm 2$ nm, and the axial crystallographic direction of the nanowire is along <110>. The gate insulator consists of a layer of HfO$_2$ deposited by atomic layer deposition, a SiO$_2$ native oxide layer, and the depleted silicon shell layer. Any doping of the silicon shell layer would simply shift the threshold voltage of the device, but threshold voltage differences are removed by the analysis procedure. The thickness of each insulator shell in this simulated device is taken to be 4 nm, 1 nm, and 1.7nm for HfO$_2$ ($\kappa$=23), SiO$_2$ ($\kappa$=3.9), and Si ($\kappa$=11.9) shell, respectively, as determined from the fabrication process.

The nanowire sits on a 50 nm thick layer of SiO$_2$ on top of an n-type silicon wafer doped with resistivity less than 0.005 Ω-cm. Using these numbers, we obtain a gate insulator capacitance as $6.9 < C_G < 10.7$ pF/cm. The device to be analyzed has a channel length of 190 nm. For details of the fabrication process and device structure, see Ref. 1.

Because of uncertainties in threshold voltage caused by charge at the dielectric/semiconductor interface and the workfunctions of different gate electrodes, it is not advisable to compare *I-V* characteristics of devices directly. It is preferable to compare $I_{ON}$ vs. $I_{ON}/I_{OFF}$ at a fixed drain voltage[18]. To generate such a curve from measurements or theoretical calculations, the device operation voltage, $V_{DD}$, is first specified. For this analysis, we take $V_{DD} = 1$ V. From the measured or calculated drain current as function of $V_{GS}$ for $V_{DS} = V_{DD}$, we extract $I_{ON}$ vs. $I_{ON}/I_{OFF}$ by defining a "window" $V_{DD}$ volts wide and superimposing it on the *I-V* characteristic. We then read $I_{ON}$ from the left side of the window and $I_{OFF}$ from the right side. By sweeping the window across the entire *I-V* characteristic, we produce a plot of $I_{ON}$ vs. $I_{ON}/I_{OFF}$.

Figure 3(a) shows the experimentally measured $I_{DS}$ vs. $V_G$ at $V_D = 1$, 0.1 and 0.01 V. The subthreshold swing at $V_D = 1$ V is 100 mV/dec. The DIBL is obtained as 150 mV/V from the horizontal displacement in the $V_D = 1$ and 0.1 V curves at $I_D = 0.1$ μA for $V_D = 1$ and 0.1 V. The series resistance of this device can be obtained from the experimental $I_D$ vs. $V_G$ characteristic at a low drain bias of $V_D = 100$ or 10 mV. The inset of Figure 3(a) shows the measured device resistance, $R_{SD} = V_{DS}/I_{DS}$ as function of $-V_G$ for $V_D = 0.1$ V and 10 mV. The saturation value of $R_{SD}$ = 5.6 k$\Omega$ appears between $V_G$ = 0 to $V_G$ = -1 V, and is the series resistance of the device.

The ambipolar behavior displayed in Fig. 3(a) raises the possibility that this device is a Schottky barrier FET. The top-of-the-barrier models we use to analyze the data assumes MOSFET-type operation in which the source can supply any current that the gate demands. Using an approach proposed by J. Appenzeller, et al.[19,20], we estimate that the barrier height of this device is only 30 meV, which may be small enough to ensure MOSFET-type operation.

Figure 3(b) compares the simulated $I_{ON}$ vs. $I_{ON}/I_{OFF}$ for two ballistic Ge NW MOSFETs with a series resistance of 5.6 K$\Omega$ as compared to the experimental measurements (triangles). Due to uncertainties of the experimental device structure, two cases were considered. The solid line presents the maximum estimated ballistic I-V of a Ge NW MOSFET with a perfectly cylindrical gate ($C_G$=10.7 pF/cm) and the largest estimated diameter with D=17 nm whereas the dashed line presents the minimum possible ballistic I-V of a Ge NW MOSFET with half-cylindrical metal gate ($C_G$=6.9 pF/cm) and the smallest possible diameter with D=13nm. Our analysis shows that this nanowire transistor operates between 60% and ~85% of the ballistic limit at $I_{ON}/I_{OFF}$=100. Even with the most conservative assumption of a cylindrical gate, the device appears to operate rather close to the ballistic limit, while the more reasonable assumption of a half cylindrical gate suggests that this device operates at its ballistic limit.

To gain further insight into the performance of this device, the number of modes involved in carrier transport, i.e., the number of modes between $E_{fs}$ and $E_{fs}$-$qV_D$ vs. gate bias was studied for two different nanowire diameters. The outside shells, Si, SiO$_2$ and HfO$_2$, remain constant (same geometry as described above), and the Fermi level is set to 100 meV above the first valence subband at equilibrium for all cases in this simulation. Figure 4 shows the number of populated modes vs. $V_G$ for a small diameter (3 nm) and a large diameter (17 nm) Ge nanowire MOSFET with a perfectly cylindrical gate under $V_{DS}$=$1$ V. We found that the number of populated modes strongly depends on the nanowire diameter and decreases as the diameter shrinks because of the larger subband separation in wires with smaller diameter. According to the geometry of the measured device, the maximum and minimum possible number of subbands dropping into the window of $E_{fs}$ and $E_{fs}$-$qV_D$ at ON-current condition for $I_{ON}$ / $I_{OFF}$ = 100 was determined to be 18 modes and 14 modes for the 17 nm diameter nanowire with a perfectly cylindrical gate and the 13 nm diameter with a half cylindrical gate, respectively. The results show that although conduction in this NWFET is not one-dimensional, a relatively small number of modes carry the current.

To observe true 1D transport at room temperature, the diameter of Ge nanowire MOSFET should be relatively small—such as 3 nm as shown in Fig. 4. For a large diameter (17 nm) nanowire MOSFETs, the number of populated modes increases quickly as soon as the device turns on. In the 3 nm case, however, the number of populated modes increases slowly as $V_G$ increases which provides a larger voltage margin to observe true 1D single-subband carrier conduction compared with larger diameter nanowires.

Our analysis of this experiment is based on a number of assumptions that should be carefully examined. First, there is the uncertainty in the precise gate capacitance, which can only be resolved by directly measuring the gate capacitance on the device under test. Second is our assumption of bulk atomic positions for the nanowire. Crystalline Si has a 4% lattice mismatch with Ge. Therefore large amount of strain might exist at the epitaxial core/shell interface. Although our preliminary examination of relaxation effects using NEMO-3D[22,23] and a similar Ge/Si nanowire core/shell structure FET (but with <100> orientation) suggests that the ON-current does not vary substantially with introduction of interfacial stress, a more extensive study of strain relaxation on carrier transport resulting from the lattice mismatch in Ge/Si core/shell nanowire MOSFETs is underway. Finally, we have not treated the self-consistent band-bending within the nanowire itself, which could also affect the electronic structure. Each of these assumptions is being examined, but we do not expect the broad conclusions to change - this device appears to operate relatively close to the ballistic limit. In spite of the uncertainties, this result is surprising, given that a recent theoretical analysis suggests that silicon nanowire FETs should operate very far from the ballistic limit when the channel length is longer than only a few nanometers[21]. Additional experimental and theoretical investigations are needed to further address the nature of room temperature high-field transport in nanowire transistors, specifically how they can operate so close to the ballistic limit.

In summary, recent experimental results for a Ge/Si core/shell NWFET device were analyzed, and the results suggest that this device operates surprisingly close to its ballistic limit. The device appears to operate as a nanowire MOSFET, with only about 14-18 modes involved in carrier transport. Our analysis also suggests that to obtain true one dimensional transport in a <110> Ge NW MOSFET, the nanowire diameter would have to be much less than about 5 nm, and the device bias would have to be carefully selected. More precise analyses of experimental data like this will require careful measurements of nanowire gate capacitance and an understanding how lattice strain and self-consistent electrostatics affect the electronic structure of nanowires. In general, the methodology presented here should prove useful for analyzing and comparing on common basis nanowire transistors of various materials and structures.

Acknowledgement

This work at Purdue was supported by the MARCO Focus Center on Materials, Structures, and Devices. Work at Harvard was supported by Defense Advanced Research Projects Agency and Intel. The authors would like to thank Raseong Kim, Kurtis Cantley, Sayeed Salahuddin, and Diego Kienle for helpful discussions.

Figure Captions:

Figure 1: (a) and (b) Electronic structure of the $D$ = 5 and 15 nm <110> cylindrical Ge nanowire, respectively. (c) Bandgap $E_G$ as function of diameter for a cylindrical <110> Ge-nanowire where $E_G$ is taken at the $\Gamma$-point of the 1D Brillouin zone. When the diameter of the nanowire is less than 5 nm, the bandgap shows a significant increase.

Figure 2: Schematics of a cylindrical gate nanowire (a) and a semi cylindrical gate nanowire sitting on the $SiO_2$ substrate (b).

Figure 3: (a) Experimental measurements of a <110> Si/Ge core/shell nanowire MOSFET: $I_{DS}$ as a function of $V_G$ at $V_D$ =1, 0.1, and 0.01 V. The vertical dashed lines show the $V_{DD}$ window used to obtain the dependence of $I_{ON}$ on $I_{ON}/I_{OFF}$ in Fig. 3(b). Based on the data of $V_D$ = 1 V, $S$ = 100 mV/decade can be determined. Using the data of $V_D$ = 0.1 and 0.01 V, the dependence of $R_{SD}$ on $V_D$ can be calculated (inset). The flat region of $R_{SD}$ can be attributed to the series resistance (at $V_G$=0, $R_{SD}$= 5.6 KΩ). (b) Simulated ON-current vs. $I_{ON}/I_{OFF}$ for a <110> Ge NW MOSFET with $R_{SD}$= 5.6 KΩ in comparison with experimental data (triangles). Solid and dashed lines present the simulated ballistic results with $D$= 17 nm and $C_G$= 10.7 pF/cm, and $D$= 13 nm and $C_G$= 6.9 pF/cm, respectively.

Figure 4: Dependence of the number of modes involved in carrier transport on the gate voltage for the 3 nm (dashed) and 17 nm (solid) diameter Ge nanowires under $V_{DS}=1$ V. The outside shells, Si, $SiO_2$ and $HfO_2$, remain constant, and the Fermi level is set to 100 meV above the first valence subband for both cases in this simulation. Compared to a large diameter nanowire such as 17 nm diameter one, the small diameter nanowire (3 nm) is relatively feasible to observe true 1D transport at room temperature due to the larger subband separation.

Figure 1

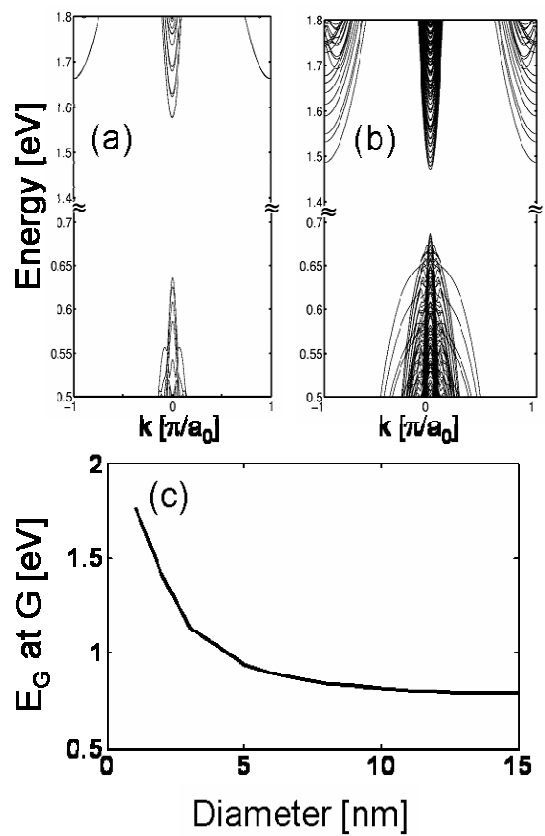

Figure 2

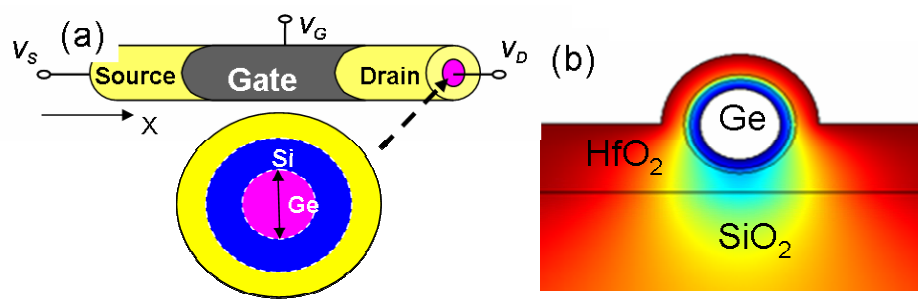

Figure 3

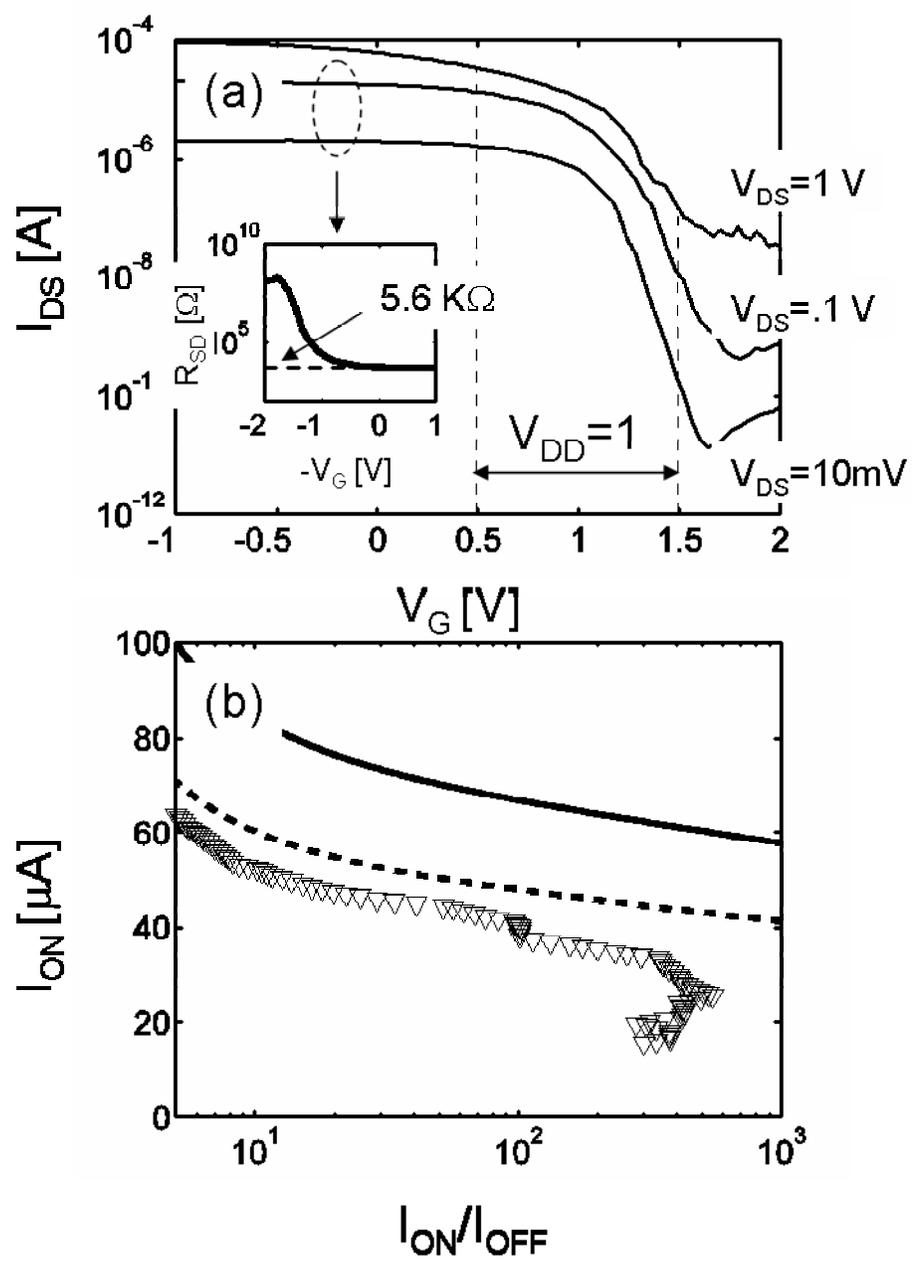

Figure 4

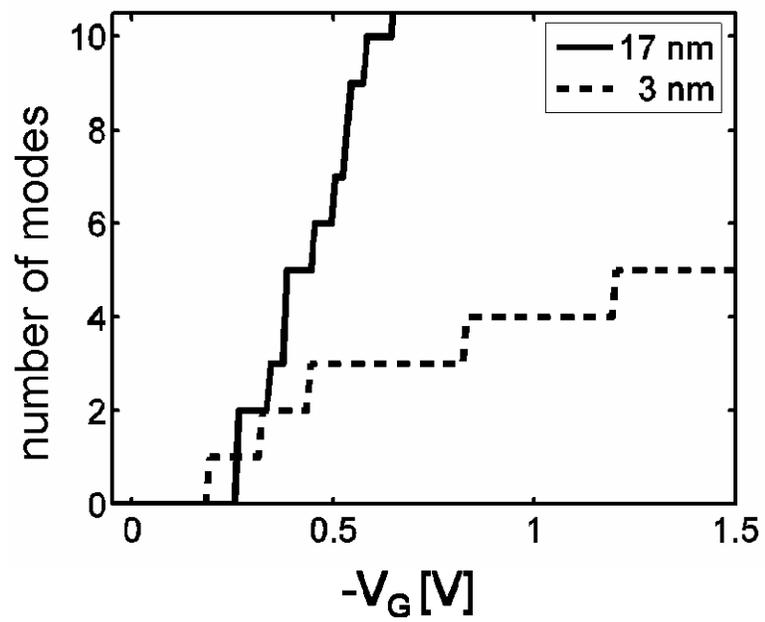